\documentclass[
prd,aps,
showpacs,tightenlines,nofootinbib,preprintnumbers, superscriptaddress
]{revtex4}
\usepackage{graphicx}
\usepackage{amssymb,amsmath,latexsym}
\usepackage{amsfonts}

\usepackage{cancel}
\usepackage{color}
\input{colordvi.tex}
\input{epsf}
\usepackage{epsf}
\usepackage{graphicx,epsfig}
\usepackage{bm}
\usepackage{latexsym,float}

\newcommand{\beq}{\begin{equation}}  \newcommand{\eeq}{\end{equation}}
\newcommand{\bef}{\begin{figure}}  \newcommand{\eef}{\end{figure}}
\newcommand{\bec}{\begin{center}}  \newcommand{\eec}{\end{center}}

\newcommand{\beqa}{\begin{eqnarray}}
\newcommand{\eeqa}{\end{eqnarray}}
\newcommand{\p}{\phi}
\newcommand{\simg}{\gtrsim}
\newcommand{\siml}{\lesssim}
\newcommand{\ka}{\kappa}

\newcommand{\Ompp}{\Omega_{\phi 0}}

\newcommand {\ga} {\ {\raise-.5ex\hbox{$\buildrel>\over\sim$}}\ }
\newcommand {\la} {\ {\raise-.5ex\hbox{$\buildrel<\over\sim$}}\ }

\def\be{\begin{equation}}
\def\ee{\end{equation}}
\def\ba{\begin{eqnarray}}
\def\ea{\end{eqnarray}}

\begin{document}

\title{Slow-roll Extended Quintessence}

\author{Takeshi Chiba}
\affiliation{Department of Physics, \\
College of Humanities and Sciences, \\
Nihon University, \\
Tokyo 156-8550, Japan}
\author{Masaru Siino}%
\affiliation{Department of Physics, 
Tokyo Institute of Technology, Tokyo 152-8551, Japan}
\author{Masahide Yamaguchi\footnote{Present address: Department of Physics, 
Tokyo Institute of Technology, Tokyo 152-8551, Japan}}%
\affiliation{Department of Physics and Mathematics, Aoyama Gakuin
University, Sagamihara 229-8558, Japan}

\date{\today}


\begin{abstract}
We derive the slow-roll conditions for a non-minimally coupled scalar
field (extended quintessence) during the radiation/matter dominated era
extending our previous results for thawing quintessence.  We find that
the ratio $\ddot\phi/3H\dot\phi$ becomes constant but negative, in sharp
contrast to the ratio for the minimally coupled scalar field.  We also
find that the functional form of the equation of state of the
scalar field asymptotically approaches that of the minimally coupled
thawing quintessence.
\end{abstract}

\maketitle

\section{Introduction}

There is strong evidence that the Universe is dominated by dark energy,
and the current cosmological observations seem to be consistent with
$\Lambda$CDM.  The equations of state of dark energy, $w$, is close to
$-1$ within 10\% or less.  This implies that even if a scalar field
(dubbed "quintessence" \cite{quint}) plays the role of dark energy, it
should roll down its potential slowly because its kinetic energy density
should be much smaller than its potential. In this situation, as in the
case of inflation, it is useful to derive the slow-roll conditions for
quintessence because the dynamics of the scalar field can be
discussed only by simple conditions without having to solve its equation
of motion directly.  Quintessence models are classified according to
their motion \cite{cl}: In "thawing" models \cite{pngb,ds,chiba} the
scalar fields hardly move in the past and begin to roll down the
potential recently, while in "freezing" models the scalar fields move in  
the opposite ways and gradually slow down the motion
\cite{rp,swz,chibatrac}. We will consider the slow-roll conditions for
thawing models since there are several particle physics models for them. 
For example, massive scalar fields (like axions or moduli) before their oscillations 
move like thawing models. Moreover, polynomial potentials beyond the Planck scale field
value can be naturally realized by F-term \cite{Fterm} and D-term
\cite{Dterm} in supergravity and in superstring \cite{Superstring}, and
axion-type potentials are obtained by instanton effects \cite{qaxion}.

In our former study, one of us (TC) derived the slow-roll conditions for
a scalar field minimally coupled to gravity \cite{chiba}. It is found
that for thawing models the acceleration term, $\ddot\phi$, is never
negligible compared with the Hubble friction term, $3H\dot\phi$, if the
Universe is dominated by radiation/matter.  Moreover the ratio,
$\ddot\phi/3H\dot\phi$, becomes constant during the radiation/matter
dominated epoch \cite{chiba}:
\beqa
\frac{\ddot\phi}{3H\dot\phi}=\frac{1+w_B}{2},
\label{ratio}
\eeqa
where $w_B$ is the equation of state of radiation/matter.  So it is
intriguing to examine to what extent the relation Eq. (\ref{ratio})
holds universally.  In \cite{cds}, one of us (TC) with Dutta and
Scherrer studied the slow-roll conditions for k-essence \cite{kess} and
again found that the relation Eq. (\ref{ratio}) persists for slow-roll
k-essence in the radiation/matter dominated era since the k-essence Lagrangian 
can be Taylor-expanded for 
small kinetic energy if it is analytical and it reduces to that of canonical scalar field by
field redefinition.

In this paper, we further study the slow-roll conditions for a scalar
field {\it non-minimally coupled} to gravity (called extended
quintessence \cite{extended}). We will find again that the ratio becomes
constant but that its value is negative:
\beqa
\frac{\ddot\phi}{3H\dot\phi}=\frac{w_B-1}{2},
\nonumber
\label{ratio2}
\eeqa
being in sharp contrast to the minimally coupled scalar field
case. Therefore, the ratio makes the non-minimally coupled scalar field
distinguishable from the minimally coupled scalar field even for very small
coupling constant $\xi$.

The slow-roll extended quintessence can provide a dynamical solution to
the coincidence problem: why dark energy dominates recently, not in the
past \cite{chiba:extended2}.  Also if the scalar field has a non-minimal
coupling during inflation, such a non-minimal coupling may provide a
dynamical solution to the fine-tuning of the initial conditions of the
scalar field.  Note that the situation is not limited to quintessence
but is applied to the case when the scalar fields which are subdominant
components in the universe move slowly. Axions, curvatons, and moduli
before the oscillation can be such fields.

The paper is organized as follows: In Sec. 2, we derive the slow-roll
conditions for non-minimally coupled scalar field during the
radiation/matter dominated epoch and discuss the dynamics of the scalar
field. In Sec. 3, we derive analytic solutions for the scalar field
during the matter dominated era to examine the slow-roll behavior of the scalar
field.  Sec. 4 is devoted to summary.

\section{Extended Thawing Quintessence}

We consider the cosmological dynamics described by the action
\beq
S=\int d^4x\sqrt{-g}\left[\frac{1}{2\ka^2}R-F(\phi)R-
\frac12(\nabla \phi)^2-V(\phi)\right]+S_m.
\label{action}
\eeq
Here $\ka^2\equiv 8\pi G_{bare}$ is the bare gravitational constant,
$F(\p)$ is the non-minimal coupling and $S_m$ denotes the action of
matter (radiation and nonrelativistic particle). We note that since matter 
is universally coupled to $g_{\mu\nu}$  in the action Eq.(\ref{action}), 
this ``Jordan frame metric'' defines the lengths and times
actually measured by laboratory rods and clocks. All experimental data
will thus have their usual interpretation in this frame.

The equations of motion in a flat FRW universe model are
\beqa
\ddot\phi + &3H\dot\phi& +V'(\phi)+6F'(\phi)\left(\dot H +2H^2\right)=0,
\label{eom:phi}\\
3H^2&=&\ka^2\left(\rho_B+\frac12\dot\phi^2+U\right)
=:\ka^2(\rho_B+\rho_{\phi}) =:\ka^2 \rho_{\rm tot},
\label{hubble}\\
2\dot H&=&-\ka^2\left( \rho_B+p_B+\rho_{\p}+\dot\phi^2/2-V-2\ddot F-4H\dot F-2F(2\dot H+3H^2)\right)\\
&=:&-\ka^2\left( (1+w_B)\rho_B+\rho_{\p}+p_{\p}\right),\nonumber\\
\label{hdot}
U &:=& V+6H\left(\dot F+HF\right), \label{u}
\eeqa
where ${}'=d/d\phi$, $\rho_B$ and $p_B$ denote the background (radiation and
matter) energy density and pressure, respectively, and $w_B=p_B/\rho_B$ is the equation of state of
radiation and matter.

\subsection{Slow-roll Conditions}

We derive the slow-roll conditions for extended (thawing) quintessence
during the matter/radiation dominated epoch. Then Eq. (\ref{eom:phi})
becomes
\beqa
&& \ddot\p+3H\dot\p+V_{\rm eff}' = 0,\label{eom:phi2}\\
&& V_{\rm eff}' \equiv V'+3F'H^2(1-3w_B).
\eeqa
By ``slow-roll'', we mean that the movement of $\phi$ during one Hubble time is
much smaller than $\phi$. On the other hand, the condition that the kinetic energy density of the scalar
field is much smaller than the potential $U$ (Eq. (\ref{u})) in the
energy density of the scalar field $\rho_{\p}$ (Eq. (\ref{hubble}))
%
%
%
%
%
\beqa
\frac12\dot\p^2 \ll U,
\label{slowroll1}
\eeqa
implies that 
\beqa
\dot\p^2H^{-2}\ll \ka^{-2} \siml \p^2, 
\eeqa
{}from $U\ll \rho_{\rm tot}\simeq \ka^{-2}H^2$ if $\ka\p\simg 1$. 
Hence we regard Eq. (\ref{slowroll1}) as the slow-roll condition.

\begin{figure}
\includegraphics[width=13cm]{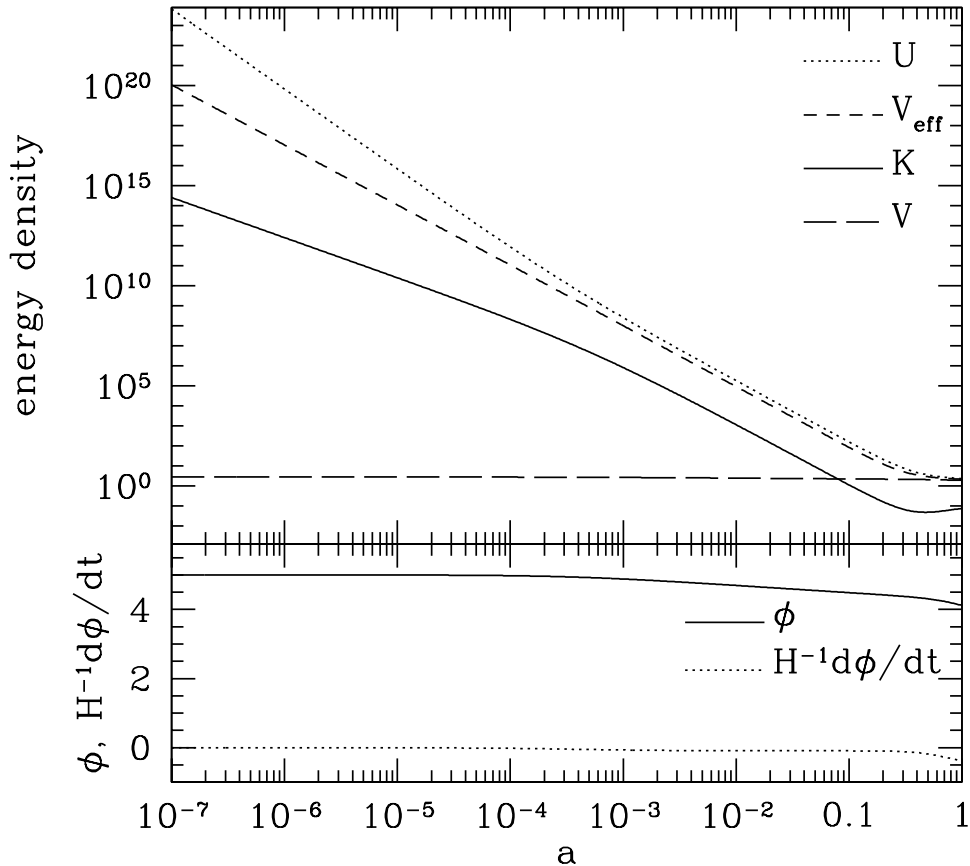}
\caption{ $K=\dot{\p}^2/2$, $V$, $V_{\rm eff}$, and $U$ (upper figure) 
and $\ka\p$ and $\ka\dot{\p}/H$ (lower figure) are shown as a
function of $a$ for a massive scalar field model with $F=\frac12 \xi
\p^2$ with $\xi=10^{-2}$. 
}
\label{VK}
\end{figure}

Note that, since the term $|3F'H^2(1-3w_B)|$ is much larger than $|V'|$
in $V_{\rm eff}'$ during the matter or the radiation dominated
era, the dynamics of $\p$ is governed by the term $3F'H^2(1-3w_B)$,
which is of the same order as $U'$. Thus, different from the minimal
coupling case ($\xi=0$), the kinetic energy density of the scalar field
is much larger than $V$
\beqa
\frac12\dot\p^2 \gg V,
\label{slowroll2}
\eeqa
during the slow-roll in the matter or the radiation dominated era. As
shown in Fig. \ref{VK}, 
during the matter or the radiation dominated era, it is realized that $V \ll \frac12\dot\p^2 \ll
U$ and also $|\dot{\p} H^{-1}| \ll \phi$, which guarantees the slow-roll
of the scalar field $\p$.

On the other hand, in the case that the scalar field dominates the
energy density of the universe, $U$ almost reduces to $V$ so that the
 condition Eq. (\ref{slowroll1}) coincides with the slow-roll condition 
defined in
Refs.~\cite{Chiba:2008ia}, in which the slow-roll conditions are
discussed in the context of the inflation with a non-minimally coupled
scalar field.

Unlike the case of inflation, $H$ is not determined by the potential
alone, but by the matter/radiation along with the scalar field energy
density so that the Hubble friction is not effective and hence $\ddot\p$
is not necessarily small compared with $3H\dot\p$ in Eq. (\ref{eom:phi})
and cannot be neglected.

Now we develop the consistent set of the slow-roll conditions.
Following \cite{cmp,chiba}, we consider the ratio,
\beqa
\beta=\frac{\ddot\p}{3H\dot\p}.
\label{beta}
\eeqa
For slow-roll (thawing) models, we first assume that $\beta$ is an ${\cal O}(1)$ 
approximately constant quantity not equal to $-1$ in the
sense $|\dot\beta|\ll H|\beta|$, and the consistency of the assumption
will be checked later. In terms of $\beta$, using Eq. (\ref{eom:phi2}),
$\dot\p$ is rewritten as
\beqa
\dot\p =-\frac{V_{\rm eff}'}{3(1+\beta)H},
\label{pdot}
\eeqa
and the condition Eq. (\ref{slowroll1}) gives the first one of the slow-roll conditions
\beqa
\epsilon:=\frac{V_{\rm eff}'^2}{6H^2U}\ll 1,
\label{slow:cond:1}
\eeqa
where we have omitted $1+\beta$ since it is an ${\cal O}(1)$ quantity
and introduced the factor of $1/6$ in $\epsilon$ \cite{chiba} so that
$\epsilon$ coincides with the inflationary slow-roll parameter,
$\epsilon=\frac12\left(\frac{V'}{\ka V}\right)^2$, if the scalar field
dominates the expansion: $H^2\simeq \ka^2V/3$ and $U \simeq V_{\rm
eff} \simeq V$.

Similar to the case of inflation, the consistency of Eq. (\ref{beta})
and Eq. (\ref{eom:phi2}) should give the second slow-roll condition. In
fact, from the time derivative of Eq. (\ref{pdot})
\beqa
\ddot\p
&=&-\frac{\dot H}{H}\dot\p-\frac{V''}{3(1+\beta)H}\dot\p-
\frac{F''H(1-3w_B)}{1+\beta}\dot\p+\frac{3F'H^2(1-3w_B)}{1+\beta}-
\frac{\dot\beta}{1+\beta}\dot\p ,
\eeqa
where we have used $(H^2(1-3w_B))^.\simeq -3H^3(1-3w_B)$.  On the other
hand, from Eq. (\ref{beta}) and Eq. (\ref{pdot}), $\ddot\p=3\beta
H\dot\p=-\beta V_{\rm eff}'/(1+\beta)$, and so we obtain
\beqa
\beta =\frac{\ddot\p}{3H\dot\p}&\simeq &
-\frac{\dot H}{3H^2}-\frac{V''}{9(1+\beta)H^2}-
\frac{F''(1-3w_B)}{3(1+\beta)}-\frac{V_{\rm eff}'-V'}{V_{\rm eff}'},\nonumber\\
&=&\frac{w_B-1}{2}-\frac{V''}{9(1+\beta)H^2}-
\frac{F''(1-3w_B)}{3(1+\beta)}+\frac{V'}{V_{\rm eff}'},
\label{betaeq}
\eeqa
where we have used $3F'H^2(1-3w_B)=V_{\rm eff}'-V'$ and $|\dot\beta|\ll
H|\beta|$. While the left-hand-side of Eq. (\ref{betaeq}) is assumed to be
an almost
time-independent quantity, the terms other than the first
in the right-hand-side are time-dependent quantities in
general. Therefore the assumption is consistent if they are negligible:\footnote{
The exception is the case of $F''={\rm const}$. In this case $F''$ needs
not to be small.  For example, if $F=\frac12 \xi\p^2$, then $\beta$
satisfies $\beta=-\frac13$ during the radiation era and $\beta=-\frac12
-\frac{\xi}{3(1+\beta)}$ so that $\beta=\frac{-9+\sqrt{9-48\xi}}{12}$
during the matter era.}
\beqa
\eta:=\frac{V''}{3H^2}; ~~~~~|\eta|\ll 1 ~~~~~{\rm and}~~~~~|F''(1-3w_B)|\ll 1~~~~~{\rm and}~~~~~\Bigl{|}\frac{V'}{V_{\rm eff}'}\Bigr{|}\ll 1,
\label{slow:cond:2}
\eeqa
so that $\beta$ becomes
\beqa
\beta=\frac{w_B-1}{2}.
\label{betasol}
\eeqa
$\beta$ given by Eq. (\ref{betasol}) is consistently an ${\cal O}(1)$ constant
not equal to $-1$.  Here the factor $1/3$ is introduced
in $\eta$ \cite{chiba} so that $\eta$ coincides with the inflationary
slow-roll parameter, $\eta=\frac{V''}{\ka^2V}$, if $H^2\simeq \ka^2V/3$.
The conditions in Eq. (\ref{slow:cond:2}) are quintessence counterparts of the
inflationary slow-roll condition $\frac{|V''|}{\ka^2V}\ll 1$.

Eq. (\ref{slow:cond:1}) and Eq. (\ref{slow:cond:2}) constitute the
slow-roll conditions for extended quintessence during matter/radiation
epoch. $\beta$ (Eq. (\ref{betasol})) is negative and is quite different
from that for a minimally coupled scalar field (Eq. (\ref{ratio})) which
is positive.  Therefore, this can be a discriminating probe of the
non-minimal coupling of the scalar field.  
Although it may be difficult to determine the thawing dynamics from distance measurements 
\cite{cds,sami}, the ratio $\beta$ may be determined 
by measuring the time variation of the fine structure constant $\alpha$
if $\phi$ induces such a variation \cite{olive} and $\alpha$ depends
linearly on $\phi$.

\begin{figure}
\includegraphics[width=13cm]{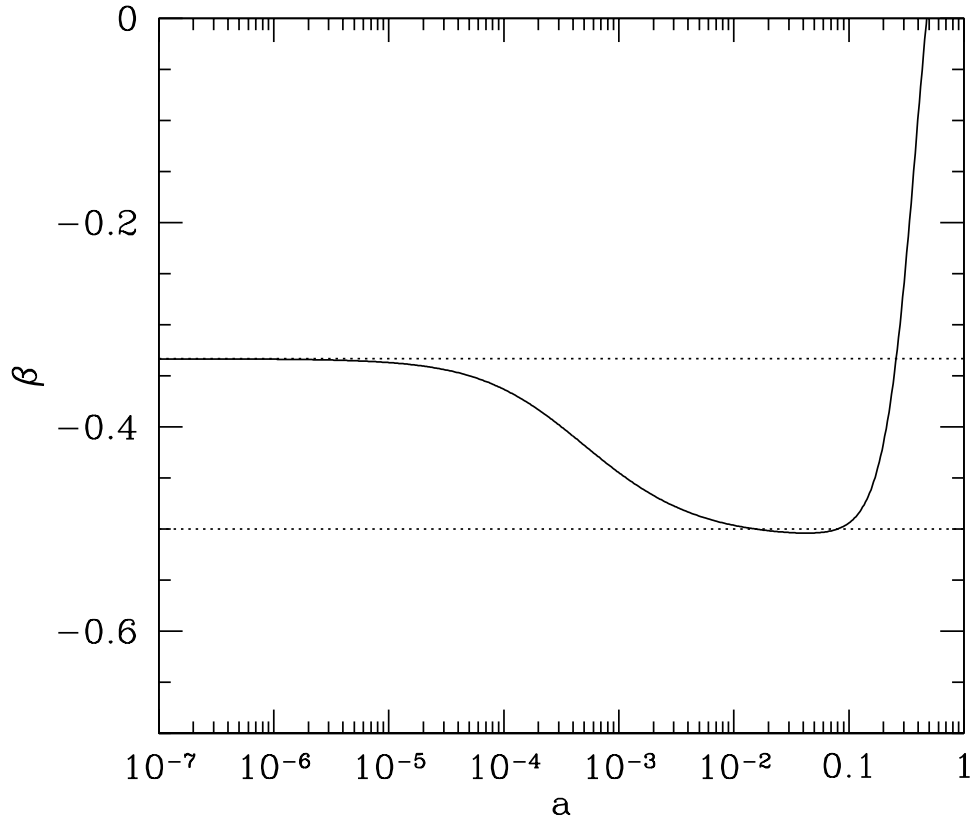}
\caption{ $\beta$ as a function of $a$ for a massive scalar field model with 
$F=\frac12 \xi \p^2$ with $\xi=10^{-2}$. The dotted lines are $\beta=-\frac13,-\frac12$, respectively. }
\label{figbeta}
\end{figure}

In Fig. \ref{figbeta}, the evolution of $\beta$ is shown for a massive
scalar field ($V=\frac12 m^2\p^2$) with a non-minimal coupling
$F=\frac12 \xi\p^2$ with $\xi=10^{-2}$. The evolution of $\beta$ agrees
nicely with Eq. (\ref{betasol}).

\subsection{Tracking without Tracking}

\label{sub:TWOT}
In the following we examine the cosmological dynamics of the extended quintessence using the slow-roll conditions Eq. (\ref{slow:cond:1}) and Eq. (\ref{slow:cond:2}).
We shall first show that the equation of state of the slow-roll extended
quintessence, $w_{\p}=p_{\p}/\rho_{\p}$, is the same as the background
equation of state, $w_B$ although the scalar field moves slowly
\cite{chiba:extended2}.

\begin{figure}
\includegraphics[width=13cm]{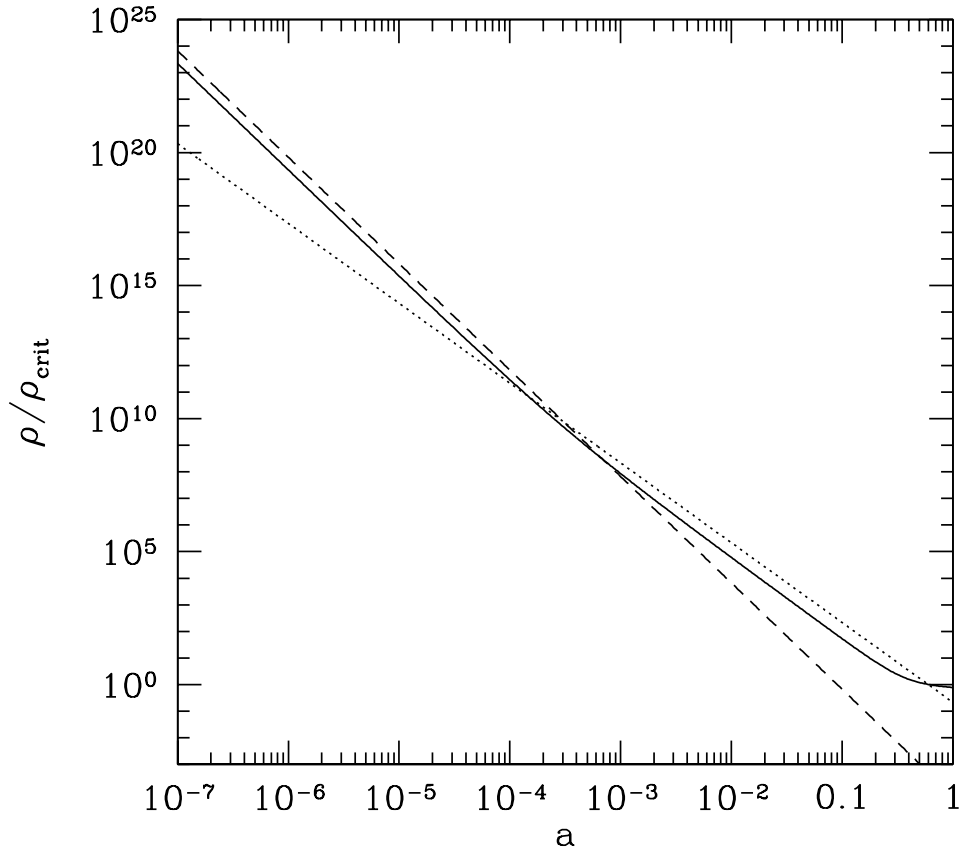}
\caption{ $\rho_{\phi}$ (solid line) as a function of $a$. 
The radiation density (dashed line) and 
the matter density (dotted line) are also shown. }
\label{figrho}
\end{figure}

Consider a slowly rolling scalar field non-minimally coupled to gravity
which satisfies the above slow-roll conditions. If the universe is
dominated by radiation/matter, the energy density of the scalar field
then becomes
\beq
\rho_{\phi}=\frac12\dot \phi^2+V +
6H\left(\dot F+HF\right)\simeq 6H^2F,
\eeq
and the pressure becomes
\beqa
p_{\phi}=\frac12 \dot\phi^2-V-2\ddot F-4H\dot F-2F(2\dot H+3H^2)
\simeq 6w_BH^2F. 
\eeqa
Since $F(\phi)$ is almost constant, this implies that the equation of
state of the scalar field $w_{\phi}=p_{\p}/\rho_{\p}$ becomes $w_B$ and
behaves as background fluid although $\phi$ itself moves hardly (See
Fig. \ref{figrho}).  We dubbed this behavior as ``tracking without
tracking''\footnote{It is no surprise that the
equation of state of tracking without tracking state is the same as
$w_B$ because in this case the Lagrangian density of the scalar field is
simply $-F(\phi)R$. Therefore the energy momentum tensor of the scalar
field is the same as the Einstein tensor which is dominated by the
background matter.}\cite{chiba:extended2}. We should emphasize that this behavior is
independent of the details of the shape of a potential as long as the
slow-roll conditions, Eq. (\ref{slow:cond:1}) and
Eq. (\ref{slow:cond:2}), are satisfied.  ``Tracking without tracking''
is rather kinematical tracker inherent in a wide class of extended
quintessence.

Next, we shall consider a scenario based on extended quintessence which
solves the coincidence problem: why dark energy becomes dominant now?
In order to ``solve'' the problem dynamically, the dark energy density
should scale in the same way as the radiation density during the
radiation dominated epoch; otherwise it is nothing but introducing a
fine-tuning to account for the coincidence from the very beginning, and
it is no surprise that there is some epoch when the two energy
components coincide.  On the other hand, however, during the matter
dominated epoch, dark energy should not track matter; otherwise dark
energy cannot dominate.

The slow-roll extended quintessence tracks the background matter during
the radiation dominated epoch (tracking without tracking) but it can
begin to move during the matter dominated epoch due to the violation of
one (or some) of the slow-roll conditions so that $w_{\p}$ deviates from
$w_B$ and the scalar field eventually dominates the universe. In
Fig. \ref{figrho}, the energy density of the scalar field is shown.

It is to be noted that one of the slow-roll conditions (the second one
in Eq. (\ref{slow:cond:2})) depends on the equation of state of
background matter $w_B$ and are automatically satisfied during the
radiation dominated epoch $w_B\simeq 1/3$.  
Hence, the non-minimal coupling with $F''\simeq {\cal{O}}(1)$ naturally realizes
the scenario.


\subsection{Dynamical Solution of Fine-tuning Problems}

We note that in order to solve the dark energy "why now problem", it is
not sufficient to explain the miniscule energy scale of dark energy, but
the solution should explain why the dark energy becomes dominant
after the matter dominated epoch. For example, particle physics models
of quintessence axion \cite{qaxion} have been constructed which realize
the tiny mass scale ($\simeq H_0\simeq 10^{-33}$ eV) via instanton
effects. However, the dynamics of the quintessence axion field is
dependent on the initial condition, and if the field is initially near
the minimum of the potential, the scalar field soon oscillates around
the minimum and can never dominate the universe.  Therefore, in the
context of quintessence axion, the why now problem is replaced with the
initial condition problem: why the scalar field started near the top of
the potential?

A non-minimal coupling can alleviate the fine-tuning.  Consider the
situation where the global PQ-like symmetry is broken (to $Z_2$) during
inflation and quintessence axion acquires a non-minimal coupling of the
form
\beqa
F(\p)=\xi f^2\cos(\p/f).
\eeqa 
Then during inflation, due to the large curvature $R\simeq 12 H^2$, the
scalar field is dynamically tuned toward the minimum of $F(\p)$.
Depending on the sign of $\xi$, the axion is thus dynamically tuned
toward the maximum/minimum of $\cos(\p/f)$: $\p\rightarrow \pi f$ for
$\xi>0$; $\p \rightarrow 0$ for $\xi<0$.  Therefore a $\xi>0$ case can
be used for a dynamical solution of the fine-tuning problem of
quintessence axion. Note that if the minimum of $F(\p)$ coincides with
the local maximum of $V(\p)$, it can hardly start rolling down and
almost behaves like the cosmological constant because the quantum
fluctuations are significantly suppressed due to the large positive
effective masses squared.

\section{Analytic Solutions}

In this section, we analytically investigate the dynamics of $\p$
and its equation of state $w_{\phi} (a)= p_{\p}/\rho_{\p}$, as done in
Refs. \cite{ds,chiba,cds}. We consider the case that the non-minimal
coupling is given by $F(\phi) = \frac12 \xi \p^2$. In this case, from
Eqs. (\ref{eom:phi}-\ref{hdot}), the equation of motion and the equation
of state are given by
\beqa
 && \ddot{\p}+3H\dot{\p}+V'(\p)+6\xi\phi\left(\dot{H}+2H^2\right)=0, \label{eom:phi3}\\   
 && w_{\phi} = p_{\p}/\rho_{\p},\\
 && \rho_{\p} = \frac12\dot{\p}^2+V
              +3\xi H\left( 2\p\dot{\p}+ H\p^2 \right), \\
 && p_{\p} = \frac12\dot{\p}^2-V-2\xi(\dot{\p}^2+\p\ddot{\p})-4\xi H \p\dot{\p}
             -\xi\p^2(2\dot{H}+3H^2).  
\eeqa
First of all, we change the variable in order to eliminate the first
derivative in Eq. (\ref{eom:phi3}) \cite{ds},
\beqa
u=(\p-\p_i) a^{3/2},
\eeqa
where $\p_i$ is an arbitrary constant and is set to the initial value
later. Then, the equation of motion becomes
\beqa
  \ddot{u}+\left[-\frac32\left(\dot{H}+\frac32 H^2\right)
                 +6\xi\left(\dot{H}+2H^2\right)\right] u
    +\left[V' +6\xi\p_i\left(\dot{H}+2H^2\right)\right] a^{\frac32}
  = 0.
\eeqa
Since we are interested in the slow-roll motion of the quintessence
field $\p$, we may expand the potential $V(\p)$ around the initial value
$\p_i$ up to the quadratic order \cite{chiba},
\beqa
  V(\p)=V(\p_i)+V'(\p_i)(\p-\p_i)+\frac12V''(\p_i)(\p-\p_i)^2.
\eeqa
Since the present equation of state of $\p$ should be almost $-1$, we
also assume that the scale factor $a(t)$ is well approximated by that in
the ${\rm \Lambda}$CDM model, which is given by \beqa
a(t)=\left(\frac{1-\Ompp}{\Ompp}\right)^{1/3}\sinh
^{2/3}(t/t_{\Lambda}), \label{scale} \eeqa
where $\Ompp$ is the present density parameter of quintessence $\p$, the
scale factor $a$ is normalized to $a=1$ 
at present, and $t_{\Lambda}$ is defined as
\beqa
t_{\Lambda} \equiv \frac{2}{\sqrt{3\ka^2 V(\p_i)}}.
\eeqa
Then, the equation motion can be written as
\beqa
 \lefteqn{\ddot{u}+\left[V''(\p_i)-\frac34 \ka^2 V(\p_i) +\xi \ka^2V(\p_i)
                 \left\{ 3+\coth^2 \left(\frac{t}{t_{\Lambda}} \right)   
                 \right\} \right] u} \nonumber \\
  &&  +\left[V'(\p_i) +\xi \ka^2 \p_i V(\p_i)
                 \left\{ 3+\coth^2\left(\frac{t}{t_{\Lambda}}\right)  
                 \right\} \right] 
     \left(\frac{1-\Ompp}{\Ompp}\right)^{1/2} 
           \sinh\left(\frac{t}{t_{\Lambda}}\right)   
  = 0.
\eeqa
Unfortunately, unlike the case of the minimal
coupling $\xi = 0$, 
we cannot solve this equation analytically for the whole
range of the cosmic time. Instead, we consider the two extreme regions. \\

\begin{itemize}

 \item Region I ($t \ll t_{\Lambda}$):

The first one is the region with $t \ll t_{\Lambda}$. In this region,
the equation of motion reduces to
\beqa
 \ddot{u}+ \frac43 \xi \frac{u}{t^2} +
     \frac43 \left(\frac{1-\Ompp}{\Ompp}\right)^{1/2} \xi \p_i
     \frac{1}{t_{\Lambda} t}  = 0,
\eeqa
whose solution is given by
\beqa
  \p(t) = \frac{1}{A}
          \left[ \left(\frac{1+A}{2}\p_i+t_i(\dot{\p})_i\right) 
                 \left(\frac{t}{t_i} \right)^{\frac{-1+A}{2}}    
               + \left(\frac{-1+A}{2}\p_i-t_i(\dot{\p})_i\right) 
                 \left(\frac{t}{t_i} \right)^{\frac{-1-A}{2}}    
          \right]
\eeqa
with $A = \sqrt{1-\frac{16}{3}\xi}$. This solution satisfies $\p=\p_i$
and $\dot{\p}=(\dot{\p})_i$ at $t=t_i$.

As was explicitly shown for the general coupling $F(\p)$ in the
subsection \ref{sub:TWOT}, in this region, the equation of state of $\p$
is almost equal to the equation of state of the background matter, that
is, $w_{\p} \simeq w_B$.\\

 \item Region II ($t \gtrsim t_{\Lambda}$):

Next, we consider the region with $t \gtrsim t_{\Lambda}$. Then, the
equation of motion reduces to
\beqa
 \ddot{u} - k^2 u
         +\left[V'(\p_i) +4 \xi \ka^2\p_i V(\p_i)
               \right] 
     \left(\frac{1-\Ompp}{\Ompp}\right)^{1/2} 
           \sinh\left(\frac{t}{t_{\Lambda}}\right)   
  = 0
\eeqa
with $k \equiv \sqrt{\left(\frac34-4\xi \right) \ka^2V(\p_i)-V''(\p_i)}$.

Apart from the coefficients, this equation coincides with that in 
the minimal coupling \cite{chiba}. Then, the solution for $K \equiv k
t_{\Lambda} \ne 1$ is given by
\beqa
 \p(t)-\p_i &=& \frac{\sinh(t_i/t_{\Lambda})}{k t_{\Lambda} \sinh(t/t_{\Lambda})}
             \left[ \sinh(kt) \cosh(k t_i)
              \left\{\frac{V'(\p_i)+4\xi\ka^2\p_iV(\p_i)}{V''(\p_i)}
              \left( \coth\left(\frac{t_i}{t_{\Lambda}}\right)
                    -k t_{\Lambda} \tanh(k t_i) \right)
              +t_{\Lambda} \dot{\p}_i \right\} \right. \nonumber \\
            && \qquad \qquad \qquad \left.     -\cosh(kt) \sinh(k t_i)
              \left\{\frac{V'(\p_i)+4\xi\ka^2\p_iV(\p_i)}{V''(\p_i)}
              \left(\coth\left(\frac{t_i}{t_{\Lambda}}\right)
                     -k t_{\Lambda} \coth(k t_i) \right)
              +t_{\Lambda} \dot{\p}_i \right\} \right] \nonumber \\ 
            &&  -\frac{V'(\p_i)+4\xi\ka^2\p_iV(\p_i)}{V''(\p_i)}.
\eeqa
As far as $a(t_{\Lambda})\ll a(t)$ and $a(t_i) \ll a(t)$, this solution may be approximated 
by that with $t_i=0=t_{\Lambda}$,
\beqa
  \p(t) = \p_i + \frac{V'(\p_i)+4\xi\ka^2\p_iV(\p_i)}{V''(\p_i)} 
           \left[ \frac{\sinh(kt)\cosh\left(\frac{t_i}{t_{\Lambda}}\right)}{k t_{\Lambda} 
                  \sinh\left(\frac{t}{t_{\Lambda}}\right)} - 1
           \right].
\eeqa

For $t \gtrsim t_{\Lambda}$, the dynamics of the field $\p$ with $\xi \ll 1$ is
determined by the potential term $V$ so that $\rho_{\p} \simeq V(\p_i)$
and $\rho_{\p}+p_{\p} = \dot{\p}^2-2\xi(\p\ddot{\p}+\dot{\p}^2)+2\xi
H\p\dot{\p}-2\xi\dot{H}\p^2 \simeq \dot{\p}^2$. Then, the
equation of state $w_{\p}$ is given by
\beqa
  1+w_{\p} &\simeq& \frac{\dot{\p}^2}{V(\p_i)} \nonumber \\
           &=& \frac{3\ka^2}{4}\cosh^2\left(\frac{t_i}{t_\Lambda}\right) \left( \frac{V'(\p_i)+4\xi\ka^2\p_iV(\p_i)}
                                      {k t_{\Lambda} V''(\p_i)}
                          \right)^2
                  \left[ \frac{k t_{\Lambda} \cosh(kt) 
                        \sinh\left( \frac{t}{t_{\Lambda}} \right)
                       - \sinh(kt)\cosh\left( \frac{t}{t_{\Lambda}} \right)}
                        {\sinh^2\left(\frac{t}{t_{\Lambda}}\right)}
                  \right]^2 \nonumber \\
           &=& (1+w_{\p0}) a^{3(K-1)} \left[
                    \frac{(K-F(a))(F(a)+1)^K+(K+F(a))(F(a)-1)^K}
                         {(K-\Ompp^{-1/2})(\Ompp^{-1/2}+1)^K
                          +(K+\Ompp^{-1/2})(\Ompp^{-1/2}-1)^K}
                                          \right]^2,
\label{eos}
\eeqa
where $w_{\p0}$ is the present equation of state of the field $\p$,
$F(a)=\sqrt{1+(\Ompp^{-1}-1)a^{-3}}$ and $K=k
t_{\Lambda}=\sqrt{1-\frac{16\xi}{3}-\frac43\frac{V''(\p_i)}{\ka^2V(\p_i)}}$. 
We have normalized the expression to $w_{\p0}$ in Eq. (\ref{eos}).  
This expression completely coincides with that in the minimal
coupling \cite{ds,chiba,cds}. However, it is noted that this expression applies only for $t
\gtrsim t_{\Lambda}$ and that the definition of $K$ is different but,  
when $\xi$ approaches $0$, reduces to that of the minimally coupled scalar field.

Finally, the solution for $K = k t_{\Lambda} = 1$ is given by
\beqa
  \p(t) = \p_i +\frac23 \left[ \frac{V'(\p_i)}{\ka^2V(\p_i)}+4\xi\p_i \right]
                \left( 1-\frac{kt}{\tanh(kt)} \right),
\eeqa
which yields the equation of state,
\beqa
  1+w_{\p} &=& \frac{\ka^2}{3} \left(\frac{V'(\p_i)+4\xi\ka^2\p_iV(\p_i)}{\ka^2V(\p_i)}
                       \right)^2
               \left( \frac{\sinh (kt)\cosh (kt)-kt}{\sinh^2(kt)}
               \right)^2  \nonumber\\
           &=& (1+w_{\p0}) \left[
               \frac{F(a)-\frac{1-\Ompp}{\Ompp a^3}
                     \ln \left\{ \sqrt{{ \frac{\Ompp a^3}{1-\Ompp}}}(1+F(a))\right\}}
                    {\Ompp^{-1/2}-\frac{1-\Ompp}{\Ompp}
                     \ln \left\{ \frac{1+\Ompp^{1/2}}{\sqrt{1-\Ompp}}\right\}}
                            \right]^2.
\eeqa

\end{itemize}

\begin{figure}
\includegraphics[width=13cm]{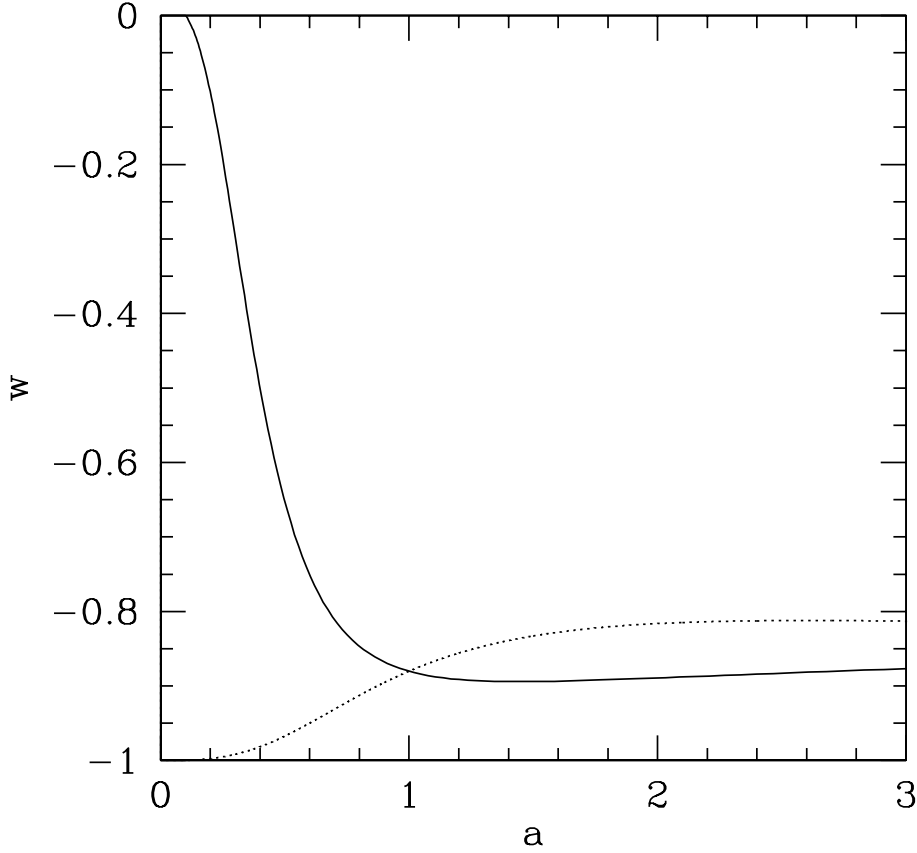}
\caption{ $w_{\p}$ as a function of $a$. The solid line is the numerical solution, while the 
dotted line is the asymptotic solution Eq. (\ref{eos}). }
\label{figeos}
\end{figure}

In Fig. \ref{figeos}, $w_{\p}$ is shown as a function of $a$.  We find
that apart from the slight offset $w_{\p}$  approaches the
asymptotic solution given by Eq. (\ref{eos}).  This, together with
\cite{cds}, makes the functional form of $w_{\p}(a)$ derived in
\cite{ds,chiba} even more useful. 
It is noted, however, that the asymptotic solution is actually a transient solution 
since the scalar field would oscillate around the minimum of $V$ in the future 
and $w_{\p}$ would tend to $0$.

\section{Summary}

We have derived the slow-roll conditions for non-minimally coupled
scalar field during the radiation/matter dominated epoch by extending
the previous results for a minimally coupled scalar field \cite{chiba}
and for non-minimally coupled inflaton(s) \cite{Chiba:2008ia}.  We have
also derived the slow-roll equation of motion of the scalar field and
found that the ratio $\ddot\phi/3H\dot\phi$ becomes constant but
negative, in sharp contrast to the result for the minimally coupled
scalar field.  This ratio can be a discriminating probe of the
non-minimal coupling of the scalar field.

We have presented two applications of the slow-roll extended
quintessence: a dynamical solution to the coincidence problem ("tracking
without tracking") and a dynamical solution to the fine-tuning problem
of quintessence axion.

We have solved the equation of motion for two limiting cases and found
that for $t\simg t_{\Lambda}$ the functional form of the equation of 
state of the scalar field
coincides with that of the minimally coupled thawing quintessence derived in
\cite{ds,chiba}. While this strengthens the universality of the
functional form of $w_{\p}(a)$ derived in \cite{ds,chiba}, this also
implies "the attraction toward minimality": the scalar field dynamics 
reduces to that of the minimally coupled scalar field. 
It would be interesting to investigate whether this property holds more generally.

\section*{Acknowledgments}

This work was supported in part by a Grant-in-Aid for Scientific
Research from JSPS (No.\,20540280(TC) and No.\,21740187(MY)) and from
MEXT (No.\,20040006(TC)) and in part by Nihon University. Some of the
numerical computations were performed at YITP at Kyoto University.




\end{document}